# Strongly correlated bistable sublattice and temperature hysteresis of elastic and thermal crystal properties

A. P. Saĭko*) and V. E. Gusakov

*Institute of Solid-State Physics and Semiconductors, National Academy of Sciences, 220072 Minsk, Belarus*

It is shown that in crystal lattices with a basis the cooperative behavior of a certain type of atoms performing optical long-wavelength vibrations in a double-well potential of the field of the matrix lattice may lead to the formation of a bistable sublattice. As a result of the interaction of the metastable states of such a sublattice with the vibrational states of the matrix lattice, the elastic and thermal properties of the crystal acquire anomalous, hysteresis-like, temperature curves. The concepts developed in the paper make it possible to obtain a qualitative interpretation, which agrees with the experimental data, of the hysteresis-like temperature dependence of the speed and absorption of ultrasonic waves, the specific heat, and the thermal conductivity in superconducting yttrium and bismuth cuprates.

## 1. INTRODUCTION

Precision experiments in which the propagation of ultrasound in high-$T_c$ superconductors[1–15] and ferroelectric conductors (see, e.g., Refs. 16 and 17) was studied detected temperature hysteresis of the speed of ultrasound (and Pal'-Val' et al.,[4] Kim et al.,[8] and Borisov et al.[17] detected temperature hysteresis of the absorption coefficient of ultrasound) that was found to encompass a temperature range from ten to hundred kelvins. More than that, in the same temperature range superconductors revealed hysteresis behavior of specific heat[18–20] and thermal conductivity.[21–25] What is remarkable is the large interval of temperature hysteresis and at the same time the absence of relaxation in the measured parameters in the hysteresis region (some samples were kept at a fixed temperature for several hours). Although there is still no universal opinion concerning the nature of the observed anomalies, it is obvious that they are related in one way or another to the metastable states of the crystal lattice. The absence of relaxation processes may indicate, for instance, that the metastable states form in conditions of a strong correlation of the lattice degrees of freedom, since otherwise local energy fluctuation would rapidly destroy the metastable states.

Below we show that this anomalous, hysteresis, behavior of the elastic and thermal characteristics of such compounds may be due to the presence in them of an anharmonically unstable, strongly correlated sublattice that executes optical long-wavelength vibrations in the field of the matrix lattice.

## 2. A MODEL OF A STRONGLY CORRELATED BISTABLE SUBLATTICE

### 2.1. Independent and strongly correlated particles in a double-well crystal potential; a bistable sublattice

In a crystal lattice with a multiatomic basis we examine a sublattice formed by ions of a single species. Suppose that each ion vibrates in a double-well asymmetric potential (Fig. 1) oriented, say, along one of the crystallographic axes. In the absence of ordering long-range forces the ions may be thought of as being almost entirely independent of one another. In this case, in addition to the intrawell vibrations in the limit $\Theta \ll U$ (where $\Theta = k_B T$ is the temperature expressed in energy units and $U$ is the height of the potential barrier), the ions are capable, due to thermal fluctuations, of performing slower movements, say, hop across the potential barrier from one stable position to another with a probability $\propto \exp\{-U/\Theta\}$. At high temperatures ($\Theta \geq U/2$) the ions are passing, i.e., they oscillate above the barrier.[26,27]

Under realistic conditions there are always correlations between the displacements of the ions in a crystal. The correlation between the relative displacement of the ions in the sublattice may be so strong that a cooperative effect could arise in which the displacement of one atom would generate similar displacements of the neighboring ions, i.e., a coherent ensemble acting as an integral whole is formed. Such a situation has a large probability of occurring in highly polarized systems. In this case a change in an external parameter, e.g., the temperature, gives rise to a coordinated shift of the atoms of the sublattice considered. The cooperative behavior of the ions of the correlated sublattice makes the sublattice unresponsive to fluctuations since, being ''bombarded'' by the quanta of the reservoir (e.g., by the phonon of the matrix lattice), the sublattice perceives a perturbation as an integral whole. Such unresponsiveness, or rigidity, of the sublattice, which prevents the separate ions from behaving independently, extends over distances of order the coherence length and means that the transition of a separate ion from intrawell dynamics to above-barrier dynamics can occur only when the entire coherent volume undergoes such a transition, since the probability of the entire correlated ensemble consisting of $n$ particles surmounting the barrier simultaneously is proportional to $\exp\{-nU/\Theta\}$ (here $U$ has the meaning of the potential of the entire sublattice per particle); for this reason, even at $n \sim 10$ the activation transitions of the correlated ensemble (and hence of each particle comprising the ensemble) across the barrier are unlikely to occur, even at temperatures $\Theta \sim U/2$.

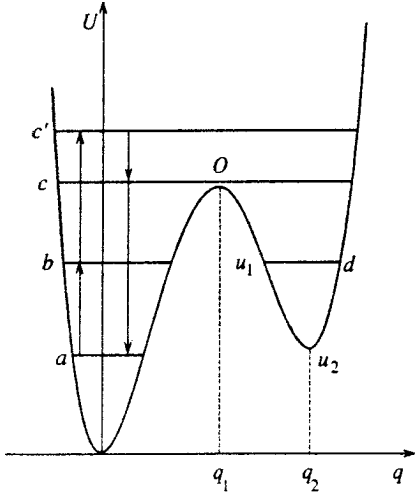

FIG. 1. The schematic of the potential and the dynamics of transitions for a strongly correlated bistable sublattice.

Thus, when being heated, a strongly correlated sublattice will evolve from vibrations in the global minimum to above-barrier vibrations, with the slower component of motion, hops from the global minimum to the local and back, being almost entirely excluded. Of course, due to renormalization, in a strongly correlated sublattice the double-well ion potential differs from the ''bare'' potential inherent in independent particles, so that it would be more natural to speak of a double-well potential for the entire coherent sublattice per particle, and all fluctuation transitions across the barrier for such a sublattice can be ignored, as we have just seen.

### 2.2. Model Hamiltonian and the derivation of the main relationships

To thoroughly study the dynamics of such a sublattice consisting of $N$ particles we need to write down the sublattice Hamiltonian $H_l$. Disregarding insignificant details, we may assume that the coherence length extends over the entire sublattice, i.e., the lattice is a single coherent ensemble. Then

$$H_l = N H_{anh}, \quad (1)$$

where $H_{anh}$ is the reduced (to a single ion) Hamiltonian of the strongly correlated lattice. We write the latter Hamiltonian in the form of an anharmonic oscillator in a double-well potential with asymmetric wells formed by the field of the matrix lattice:

$$H_{anh} = \frac{p^2}{2m} + \frac{\alpha}{2} q^2 - \frac{\beta}{3} q^3 + \frac{\gamma}{4} q^4, \quad (2)$$

where $m$ is the ion mass and $q$ and $p$ are the coordinate and canonically conjugate momentum of the ion along a specified direction fixed, say, by one of the crystallographic axes.

We examine the thermal behavior of the sublattice in the approximation of self-consistent phonons.[28] To this end we introduce the statistical-mean displacement $\langle q \rangle$, the dynamic displacement $\delta q(t) = q(t) - \langle q \rangle$, and the variance $\sigma \equiv \langle (q - \langle q \rangle)^2 \rangle$. Then, using the relationships $\langle (\delta q)^{2n} \rangle = (2n-1) \times \ldots \times 3 \sigma^n$ and $\langle (\delta q)^{2n+1} \rangle = 0$ ($n$ is an integer), valid in the adopted approximation, we find that

$$\langle H_{anh} \rangle = \frac{\alpha}{2} \langle q \rangle^2 - \frac{\beta}{3} \langle q \rangle^3 + \frac{\gamma}{4} \langle q \rangle^4 + m \Omega^2 \sigma - \frac{3}{4} \gamma \sigma^2, \quad (3)$$

where

$$\Omega^2 \equiv \frac{[\alpha - 2\beta \langle q \rangle + 3\gamma(\sigma + \langle q \rangle^2)]}{m} \quad (4)$$

is the effective frequency characterizing the sublattice. The variance $\sigma$ can be found by the fluctuation-dissipation theorem:

$$\sigma = \frac{1}{2m\Omega} \coth \frac{\Omega}{2\Theta} \quad (5)$$

(we set $\hbar = 1$ throughout the paper), the relation between $\sigma$ and $\langle q \rangle$ can be established from the condition for stability of the sublattice, $\langle \partial H_{anh} / \partial (\delta q) \rangle = 0$:

$$(\beta - 3\gamma \langle q \rangle) \sigma = \alpha \langle q \rangle - \beta \langle q \rangle^2 + \gamma \langle q \rangle^3. \quad (6)$$

The free energy $F$, which we must know in order to give a complete statistical-thermodynamic description of the system, is established by the Bogolyubov variational principle:

$$F \leq F_0 - \langle H_{anh} - H_0 \rangle_0 = \frac{\alpha}{2} \langle q \rangle^2 - \frac{\beta}{3} \langle q \rangle^3 + \frac{\gamma}{4} \langle q \rangle^4 + \Theta \ln \left( 2 \sinh \frac{\Omega}{2\Theta} \right) - \frac{3}{4} \gamma \sigma^2, \quad (7)$$

where $F_0$ is the free energy corresponding to the Hamiltonian of the pseudoharmonic approximation,

$$H_0 = \frac{\alpha}{2} \langle q \rangle^2 - \frac{\beta}{3} \langle q \rangle^3 + \frac{\gamma}{4} \langle q \rangle^4 + \frac{p^2}{2m} + \frac{m}{2} \Omega^2 (\delta q)^2. \quad (8)$$

The closed system of equations (3)–(8) makes it possible, at least qualitatively, to describe the thermal behavior of the bistable sublattice model considered here.

### 2.3. Temperature dependence of the dynamic and statistical characteristics of a bistable sublattice

The results of numerical calculations by formulas (4)–(7) are illustrated by Figs. 2 and 3.

Figure 2(a) depicts the temperature dependence of the mean displacement $\langle q \rangle$ of the sublattice. The solution represented by curve 1, which lies below the asymptote (the dashed line) and has the shape of a hysteresis curve, describes the transition of the sublattice from the global minimum (see Fig. 1) to passing trajectories (above-barrier oscillations) when the system temperature is raised from absolute zero; in the high-temperature limit, $\langle q \rangle$ approaches its asymptotic value $\langle q \rangle_{as} = \beta/3\gamma$. Above the asymptote there is the solution represented by curve 2, which is related to the possibility of the sublattice being in the second, local, minimum at low temperatures. However, the probability that the local minimum will be occupied is extremely low, since the thermal fluctuations that would take the system from the global minimum to the local one are suppressed by the strongly correlated movements of the atoms in the sublattice (see Sec. 2.1). Hence the contribution of this solution [see also Figs. 2(b) and 2(c)], which would be effective for the case of independent particles, will be ignored throughout the paper (nor will it be depicted in the figures, with the exception of Figs. 3 and 5).

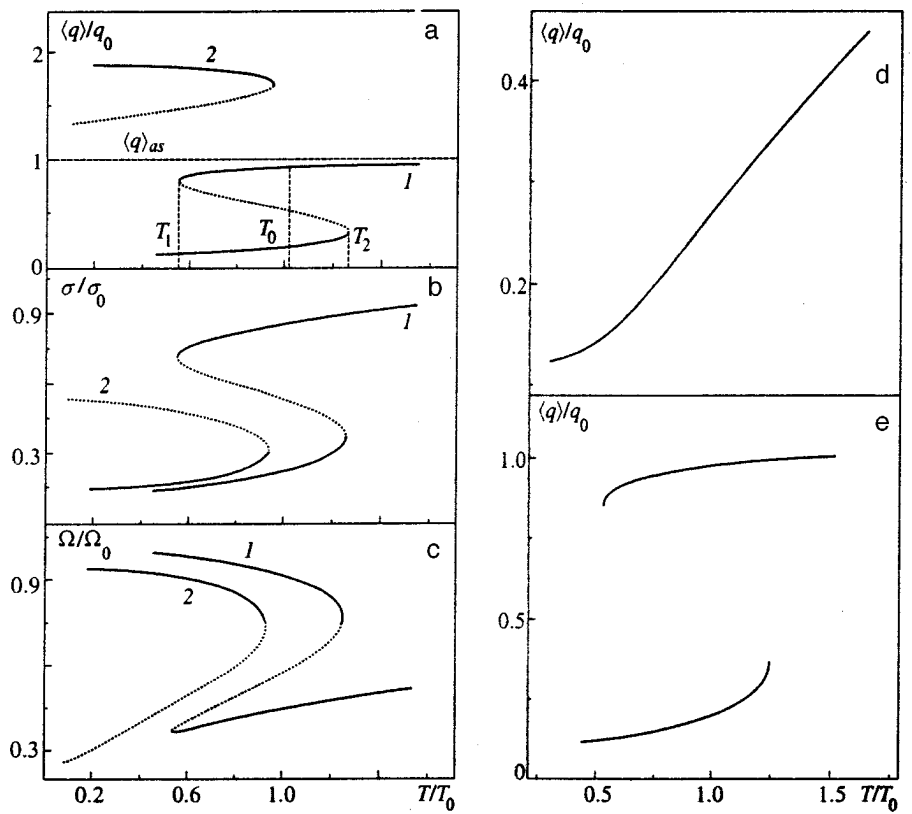

FIG. 2. Temperature dependence of the statistical-mean displacement $\langle q \rangle$ (a), the variance $\sigma$ (b), and the effective frequency $\Omega$ (c) of a bistable sublattice. (d) The statistical-mean displacement $\langle q \rangle$ for an anharmonic oscillator calculated by the molecular-dynamics method with allowance for kinetic-energy fluctuations; (e) the same for the case where the kinetic-energy fluctuations are weaker by a factor of 100; $q_0 = \langle q \rangle_{as}$, $\sigma_0 = \sigma(q_0)$, and $\Omega_0 = \Omega(q_0, \sigma_0)$. The parameters of the bistable potential are $u_1 = 0.03$ eV, $q_1 = 0.073$ Å, $q_2 = 0.14$ Å, and $T_0 = 173$ K. The curves 1 and 2 describe the motion of the lattice in the global and local minima of the potential, respectively. The dotted curves represent unstable solutions.

The hysteresis behavior of the statistical-mean displacement $\langle q \rangle$, the displacement variance $\sigma$, and the effective frequency $\Omega$ of the sublattice (Fig. 2) is explained by the nature of the temperature dependence of the free energy $F$ per particle (Fig. 3). The "low-temperature" branches of the hysteresis curves in Fig. 2 correspond to the free energy of the sublattice in the left, global, minimum (curve 1 in Fig. 3), while the "high-temperature" branches correspond to the free energy of above-barrier vibrations (curve 2 in Fig. 2); curve 3 in Fig. 3 describes unstable states; and curve 4 corresponds to solutions that refer to the positions of the sublattice in the right, local, minimum.

At the point $T_0$ the free energies become equal, and under the condition of total equilibrium at this point there would have been a transition of the sublattice from intrawell vibrations to above-barrier vibrations or back, depending on whether the system is heated or cooled. However, Fig. 3 shows (and so does Fig. 2) that "overheated" [in the interval $(T_0, T_2)$] or "supercooled" [in the interval $(T_0, T_1)$] metastable states may set in (see the discussion below). When heated, the sublattice, reaching the boundary of the metastable region at point $T_2$, suddenly changes its dynamics: it undergoes a first-order transition from intrawell vibrations to above-barrier vibrations, with the frequency decreasing approximately twofold (see Figs. 2 and 3); when cooled, the sublattice, entering the region of metastable states and approaching the region boundary at point $T_1$, discontinues its above-barrier motion and "falls" into the deeper potential well, retaining its previous frequency of vibrations in the process. The hysteresis region $\Delta T = T_2 - T_1$ depends primarily on the depth of the local minimum: as the depth decreases, $\Delta T$ becomes smaller, and at the transition point $T_c$ vanishes completely; at this point the first and second temperature derivatives of $\langle q \rangle$, $\sigma$, and $\Omega$ become infinite, which corresponds to a second-order transition.

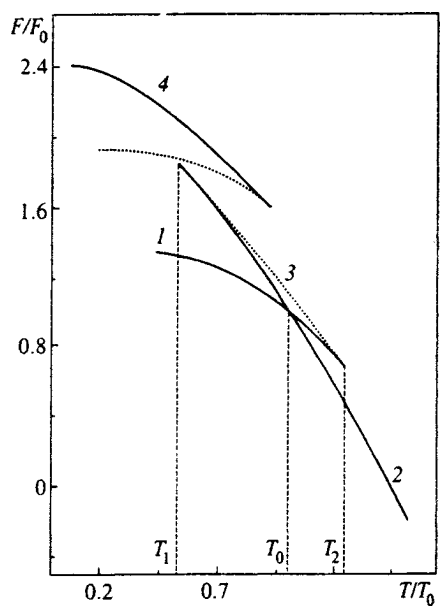

FIG. 3. Temperature dependence of the free energy of a bistable sublattice: curves 1 and 4 correspond to states in the global and local minima, curve 2 corresponds to above-barrier states, and curve 3 corresponds to an unstable solution; $u_1 = 0.03$ eV, $q_1 = 0.073$ Å, $q_2 = 0.14$ Å, and $T_0 = 173$ K.

Thus, at the points $T_1$ and $T_2$, where the boundaries of the metastable regions are crossed, at the point $T_0$ where the free energies are equal, and at the transition point $T_c$ the states of the correlated, ordered, sublattice change, i.e., order-order transitions of the first- or second orders occur.

What is interesting is that the results of molecular-dynamics modeling agree completely with our results obtained by the self-consistent phonon approximation. Molecular-dynamics calculations (Fig. 2) show that when the fluctuation hops across the barrier are effective (the case of independent particles), even at fairly moderate temperatures the second, local, minimum may become occupied (in accordance with the Boltzmann factor $\exp\{-U/\Theta\}$), and we will have a monotonic temperature dependence [Fig. 2(d)] of the displacement and hence of the other parameters, the displacement variance and the effective frequency of the system's vibrations. As the probability of fluctuation transitions drops, due to the realization of a correlated state in the sublattice, the temperature dependence of the displacement and the other parameters acquires the shape of a hysteresis curve.

### 2.4. Qualitative interpretation of the hysteresis behavior of a bistable lattice with strongly correlated particles

The hysteresis behavior of a strongly correlated sublattice in an asymmetric double-well potential can easily be understood from qualitative physical considerations based on classical statistical physics.

In view of the broken symmetry, at low temperatures the sublattice is in the left, global, minimum. As a result of heating, the oscillation trajectories (the behavior of the system is examined using the tools of classical statistical physics) of the ions of the correlated sublattice gradually rise to the vertex $O$ (Fig. 1) of the potential barrier, because the fluctuations in a coherent ensemble are suppressed (this aspect was discussed earlier). At first glance, when on path $b$, the sublattice (and hence each ion) could, as its temperature grows, either go over to the closest passing orbit $c$ by reducing its velocity (since the distance between the stopping points increases) or find itself in the right well. However, both cases are impossible. Finding itself, for instance, on the path $c$, which is directly above the barrier, and hence reducing its kinetic energy, the sublattice would be at a lower temperature (since in the classical limit the temperature is simply the average kinetic energy), and this would violate the isothermal condition (the sublattice is in contact with the thermostat). Hence eventually the sublattice will find itself on a higher path, say $c'$, starting from which it will rise higher and higher as the temperature increases. Neither can the sublattice go from path $b$ in the left well to path $d$ in the right, since it would have to lower its average kinetic energy, i.e., its temperature. We now consider the reverse course that the system takes when the temperature is lowered. As the temperature of the thermostat decreases, the sublattice gradually goes over to lower paths down to path $c$ on which the sublattice average kinetic energy $K_c$, i.e., the temperature $T_c$, is much lower than the one ($K_{c'}$ or $T_{c'}$) that the sublattice had when it emerged from the left well and went over to passing trajectories. A further decrease in temperature will force the sublattice to ''fall'' onto one of the low-lying paths in the left well (due to the broken symmetry), say onto path $a$, without changing its average kinetic energy, i.e., $K_a$ will be equal to $K_c$, or $T_a = T_c$. Thus, the size of the hysteresis region is $K_{c'} - K_c = K_b - K_a$ or, what is the same thing, $T_{c'} - T_c = T_b - T_a$.

## 3. INTERACTION BETWEEN THE BISTABLE SUBLATTICE AND THE MATRIX LATTICE

### 3.1. The total lattice Hamiltonian

The interaction of the anharmonic vibrations of a bistable lattice and the phonon excitations of the matrix lattice may lead to experimentally observable effects. For instance, the scattering of a traveling acoustic mode of the matrix lattice by perturbations caused by the vibrational motion of the bistable lattice in a double-well potential gives rise to singularities in the real and imaginary parts of these modes, which must be observable in experiments, in particular, in the anomalous behavior of the elastic and thermal characteristics of the crystal, such as the speed (the elastic modulus) and decay of ultrasound and the thermal conductivity. Thus, to study these questions we must focus on the interaction between the vibrational degrees of freedom of the matrix and the bistable sublattice.

We write the total lattice Hamiltonian $H$ normalized to the number of atoms in the bistable sublattice as

$$H = H_h + H_{\text{anh}} + H_{\text{int}}. \qquad (9)$$

The first term on the right-hand side of Eq. (9), $H_h$, models the phonon Hamiltonian of the lattice and is taken in the form of the Hamiltonian of a set of harmonic oscillators whose parameters are normalized to the empirical values of the lattice constants of the crystal:

$$H_h = \sum_k \left( \frac{p_k^2}{2\mu_k} + \frac{\mu_k \omega_k^2}{2} x_k^2 \right), \qquad (10)$$

where $x_k$, $p_k$, $\mu_k$, and $\omega_k$ are the displacement, momentum, mass, and frequency of the $k$th oscillator (mode). The term $H_{\text{anh}}$ is the Hamiltonian describing the bistable sublattice; it is defined in (2). The last term in (9), $H_{\text{int}}$, allows for the coupling of the lattice oscillators and the bistable sublattice; it is chosen in the form of a sum of the cubic and quartet interaction terms:

$$H_{\text{int}} = H_{\text{int}}^{(3)} + H_{\text{int}}^{(4)}, \qquad (11)$$

$$H_{\text{int}}^{(3)} = q^2 \sum_k \lambda_k x_k, \quad H_{\text{int}}^{(4)} = q^2 \sum_{k,k'} \lambda_{kk'} x_k x_{k'}, \qquad (12)$$

where $\lambda_k$ and $\lambda_{kk'}$ are the coupling coefficients.

### 3.2. Deriving the Dyson equation for the phonon Green's function

To find the renormalized frequencies of the matrix lattice and the corresponding decay coefficients, we use the method of equations of motion for two-time retarded Green's functions:[29]

$$\langle\langle x_k(t);x_{k'}(t')\rangle\rangle = -i\theta(t-t')\langle[x_k(t),x_{k'}(t')]\rangle$$

$$= \frac{1}{2\pi}\int_{-\infty}^{\infty}d\omega\exp\{-i\omega(t-t')\}$$

$$\times\langle\langle x_k|x_{k'}\rangle\rangle_\omega, \quad (13)$$

where $\langle\cdots\rangle$ denotes the operation of quantum-statistical averaging, $[\cdots,\cdots]$ stands for a commutator, and $\theta(\tau)$ is the Heaviside step function. Differentiating the Green's function (13) first with respect to $t$ and then with respect to $t'$ as described in Ref. 28, we arrive at three coupled equations for the Fourier transforms of the Green's functions:

$$\sum_{k''}[\mu_k(\omega^2-\omega_k^2)\delta_{kk''}-2(\sigma+\langle q\rangle^2)\lambda_{kk''}]\langle\langle x_{k''}|x_{k'}\rangle\rangle_\omega$$

$$= \delta_{kk'}+\sum_{k''}2\lambda_{kk''}\cdot{}^{\text{ir}}\langle\langle Qx_{k''}|x_{k'}\rangle\rangle_\omega+\lambda_k\langle\langle Q|x_{k'}\rangle\rangle_\omega, \quad (14)$$

$$\sum_{k_1}[\mu_{k'}(\omega^2-\omega_{k'}^2)\delta_{k'k_1}-2(\sigma+\langle q\rangle^2)\lambda_{k'k_1}]\langle\langle Q|x_{k_1}\rangle\rangle_\omega$$

$$= \sum_{k_1}2\lambda_{k'k_1}\langle\langle Q|Qx_{k_1}\rangle\rangle_\omega^{\text{ir}}+\lambda_{k'}\langle\langle Q|Q\rangle\rangle_\omega, \quad (15)$$

$$\sum_{k_1}[\mu_{k'}(\omega^2-\omega_{k'}^2)\delta_{k'k_1}-2(\sigma+\langle q\rangle^2)\lambda_{k'k_1}]\cdot{}^{\text{ir}}\langle\langle Qx_{k''}|x_{k_1}\rangle\rangle_\omega$$

$$= \sum_{k_1}2\lambda_{k'k_1}\cdot{}^{\text{ir}}\langle\langle Qx_{k''}|Qx_{k_1}\rangle\rangle_\omega^{\text{ir}}+\lambda_{k'}\cdot{}^{\text{ir}}\langle\langle Qx_{k''}|Q\rangle\rangle_\omega, \quad (16)$$

where

$$Q \equiv 2\langle q\rangle\delta q + (\delta q)^2. \quad (17)$$

The superscript "ir" indicates that the corresponding Green's function is irreducible, i.e., it cannot be reduced to lower-order functions by decoupling the product of single-time operators.[28]

We define the Green's function in the lowest-order approximation as

$$\sum_{k''}[\mu_{k'}(\omega^2-\omega_k^2)\delta_{kk''}-2\lambda_{kk''}$$

$$\times(\sigma+\langle q\rangle^2)]\langle\langle x_{k''}|x_{k'}\rangle\rangle_\omega^{(0)} = \delta_{kk'}. \quad (18)$$

Then Eqs. (14)–(16) become

$$\langle\langle x_k|x_{k'}\rangle\rangle_\omega = \langle\langle x_k|x_{k'}\rangle\rangle_\omega^{(0)} + \sum_{k'',k_1}2\lambda_{k''k_1}\langle\langle x_k|x_{k''}\rangle\rangle_\omega^{(0)}$$

$$\times{}^{\text{ir}}\langle\langle Qx_{k_1}|x_{k'}\rangle\rangle_\omega + \sum_{k''}\lambda_{k''}\langle\langle x_k|x_{k''}\rangle\rangle_\omega^{(0)}$$

$$\times\langle\langle Q|x_{k'}\rangle\rangle_\omega, \quad (19)$$

$$\langle\langle Q|x_{k'}\rangle\rangle_\omega = \sum_{k''}\lambda_{k''}\langle\langle Q|Q\rangle\rangle_\omega\cdot\langle\langle x_{k''}|x_{k'}\rangle\rangle_\omega^{(0)}$$

$$+ \sum_{k'',k_1}2\lambda_{k_1k''}\langle\langle Q|Qx_{k_1}\rangle\rangle_\omega^{\text{ir}}\cdot\langle\langle x_{k''}|x_{k'}\rangle\rangle_\omega^{(0)}, \quad (20)$$

$${}^{\text{ir}}\langle\langle Qx_{k_1}|x_{k'}\rangle\rangle_\omega = \sum_{k''}\lambda_{k''}\cdot{}^{\text{ir}}\langle\langle Qx_{k_1}|Q\rangle\rangle_\omega\cdot\langle\langle x_{k''}|x_{k'}\rangle\rangle_\omega^{(0)}$$

$$+ \sum_{k'',k_0}2\lambda_{k_0k''}\cdot{}^{\text{ir}}\langle\langle Qx_{k_1}|Qx_{k_0}\rangle\rangle_\omega^{\text{ir}}$$

$$\times\langle\langle x_{k''}|x_{k'}\rangle\rangle_\omega^{(0)}, \quad (21)$$

thus yielding

$$\langle\langle x_k|x_k\rangle\rangle_\omega = \langle\langle x_k|x_k\rangle\rangle_\omega^{(0)} + \sum_{k',k''}\lambda_{k'k''}\langle\langle x_k|x_{k'}\rangle\rangle_\omega^{(0)}$$

$$\times\langle\langle Q|Q\rangle\rangle_\omega\cdot\langle\langle x_{k''}|x_k\rangle\rangle_\omega^{(0)}$$

$$+ \sum_{k',k'',k_0,k_1}4\lambda_{k',k''}\lambda_{k_0,k_1}\langle\langle x_k|x_{k'}\rangle\rangle_\omega^{(0)}$$

$$\times{}^{\text{ir}}\langle\langle Qx_{k''}|Qx_{k_0}\rangle\rangle_\omega^{\text{ir}}\cdot\langle\langle x_{k_1}|x_k\rangle\rangle_\omega^{(0)}$$

$$+ \sum_{k',k'',k_1}2\lambda_{k'k''}\lambda_{k_1}\langle\langle x_k|x_{k'}\rangle\rangle_\omega^{(0)}$$

$$\times{}^{\text{ir}}\langle\langle Qx_{k''}|Q\rangle\rangle_\omega\cdot\langle\langle x_{k_1}|x_k\rangle\rangle_\omega^{(0)}$$

$$+ \sum_{k',k'',k_1}2\lambda_{k_1k''}\lambda_{k'}\langle\langle x_k|x_{k'}\rangle\rangle_\omega^{(0)}$$

$$\times\langle\langle Q|Qx_{k_1}\rangle\rangle_\omega^{\text{ir}}\cdot\langle\langle x_{k''}|x_k\rangle\rangle_\omega^{(0)}. \quad (22)$$

The two last "interference" terms in (22), containing the Green's functions ${}^{\text{ir}}\langle\langle Qx_k|Q\rangle\rangle$ and $\langle\langle Q|Qx_k\rangle\rangle^{\text{ir}}$, have an order of smallness in the interaction $H_{\text{int}}$ higher than the second and can be discarded. We also ignore the contribution of off-diagonal components of the Green's functions in the lowest-order approximation, assuming that $\langle\langle x_k|x_{k'}\rangle\rangle_\omega^{(0)} = \delta_{kk'}\langle\langle x_k|x_k\rangle\rangle_\omega^{(0)}$, and write Eq. (22) in the form of the Dyson equation:

$$\langle\langle x_k|x_k\rangle\rangle_\omega^{-1} = (\langle\langle x_k|x_k\rangle\rangle_\omega^{(0)})^{-1} - M_k(\omega), \quad (23)$$

where the self-energy part has the form

$$M_k(\omega) = \lambda_k^2\langle\langle Q|Q\rangle\rangle_\omega + \sum_{k',k''}4\lambda_{kk'}\lambda_{k''k}$$

$$\times{}^{\text{ir}}\langle\langle Qx_{k'}|Qx_{k''}\rangle\rangle_\omega^{\text{ir}}, \quad (24)$$

with

$$\langle\langle Q|Q\rangle\rangle_\omega = 4\langle q\rangle^2\langle\langle\delta q|\delta q\rangle\rangle_\omega + \langle\langle(\delta q)^2|(\delta q)^2\rangle\rangle_\omega$$

$$+ 2\langle q\rangle(\langle\langle\delta q|(\delta q)^2\rangle\rangle_\omega + \langle\langle(\delta q)^2|\delta q\rangle\rangle_\omega), \quad (25)$$

$$^{ir}\langle\langle Qx_k|Qx_{k'}\rangle\rangle^{ir}_\omega = 4\langle q\rangle^2 \langle\langle \delta q x_k|\delta q x_{k'}\rangle\rangle_\omega$$
$$+ {}^{ir}\langle\langle(\delta q)^2 x_k|(\delta q)^2 x_{k'}\rangle\rangle^{ir}_\omega + 2\langle q\rangle$$
$$\times ({}^{ir}\langle\langle\delta q x_k|(\delta q)^2 x_{k'}\rangle\rangle^{ir}_\omega$$
$$+ {}^{ir}\langle\langle(\delta q)^2 x_k|\delta q x_{k'}\rangle\rangle_\omega). \quad (26)$$

[it should be recalled that the operator $Q$ is defined in Eq. (17)].

### 3.3. An approximate calculation of the self-energy part

It is convenient to express the higher-order Green's functions that enter into $M_k(\omega)$ [see Eqs. (25) and (26)] in terms of correlation functions via the spectral theorem.[29] For instance, for one of these Green's functions we have

$$^{ir}\langle\langle(\delta q)^2 x_k|(\delta q)^2 x_{k'}\rangle\rangle^{ir}_\omega$$
$$= \frac{1}{2\pi}\int_{-\infty}^\infty \frac{d\omega'}{\omega-\omega'}\left(\exp\frac{\omega'}{\Theta}+1\right)\int_{-\infty}^\infty dt$$
$$\times \exp\{-i\omega' t\}\langle {}^{ir}[(\delta q(t))^2 x_k(t)][(\delta q(t'))^2 x_{k'}(t')]^{ir}\rangle. \quad (27)$$

The correlation function in (27) can be decoupled by forming pairwise two-time averages (single-time averages, according to the definition of the "ir" operation, are equal to zero):

$$\langle {}^{ir}[(\delta q(t))^2 x_k(t)][(\delta q(t'))^2 x_{k'}(t')]^{ir}\rangle$$
$$\approx 2\delta_{kk'}\langle \delta q(t)\delta q(t')\rangle^2 \cdot \langle x_k(t)x_k(t')\rangle, \quad (28)$$

where the factor 2 reflects the two possible ways of pairing the operators $\delta q$ taken at times $t$ and $t'$. By analogy with the diagrammatic technique, we can assume that adopting the approximation (28) is equivalent to ignoring the vertex corrections in the processes of interaction between phonons and the vibrations of the bistable sublattice. The spectral theorem can be used to express the one-particle correlators in (28) in terms of the corresponding Green's functions:

$$^{ir}\langle\langle(\delta q)^2 x_k|(\delta q)^2 x_{k'}\rangle\rangle^{ir}_\omega$$
$$\approx 2\delta_{kk'}\int_{-\infty}^\infty\int\int \frac{d\omega_1 d\omega_2 d\omega_3}{\omega-(\omega_1+\omega_2+\omega_3)}$$
$$\times \frac{\exp\{(\omega_1+\omega_2+\omega_3)/\Theta\}-1}{[\exp\{\omega_1/\Theta\}-1][\exp\{\omega_2/\Theta\}-1][\exp\{\omega_3/\Theta\}-1]}$$
$$\times \left[-\frac{1}{\pi}\mathrm{Im}\langle\langle\delta q|\delta q\rangle\rangle_{\omega_1+i\varepsilon}\right]\left[-\frac{1}{\pi}\mathrm{Im}\langle\langle\delta q|\delta q\rangle\rangle_{\omega_2+i\varepsilon}\right]$$
$$\times \left[-\frac{1}{\pi}\mathrm{Im}\langle\langle x_k|x_k\rangle\rangle_{\omega_3+i\varepsilon}\right]. \quad (29)$$

In the same way one should deal with the remaining Green's functions in the expression for $M_k(\omega)$; some of these, namely those in the parentheses in Eqs. (25) and (26), vanish in view of approximations of the form (28). Next, in calculating (29) we can ignore the self-energy parts of the one-particle Green's functions by writing them in the lowest-order approximation:

$$\langle\langle\delta q|\delta q\rangle\rangle_\omega \to \langle\langle\delta q|\delta q\rangle\rangle^{(0)}_\omega = [m(\omega^2-\Omega^2)]^{-1}, \quad (30)$$

with $\langle\langle\delta q|\delta q\rangle\rangle^{(0)}_\omega$ the Green's function corresponding to the Hamiltonian $H_0$ [Eq. (8)], and

$$\langle\langle x_k|x_k\rangle\rangle_\omega \to \langle\langle x_k|x_k\rangle\rangle^{(0)}_\omega$$
$$= [\mu_k(\omega^2-\omega_k^2)-2\lambda_{kk}(\sigma+\langle q\rangle^2)]^{-1}. \quad (31)$$

This enables us to explicitly calculate the exact Green's function $\langle\langle x_k|x_k\rangle\rangle_\omega$ on the basis of (23).

### 3.4. Determining the shift and decay of the lattice mode frequencies

The renormalized frequencies of the lattice modes, $\varepsilon_k$, and the decay coefficients $\Gamma_k$ can be found by solving the equation

$$(\langle\langle x_k|x_k\rangle\rangle^{(0)}_\omega)^{-1} - \mathrm{Re}\, M_k(\tilde\omega_k+i\varepsilon) + i\,\mathrm{Im}\, M_k(\tilde\omega_k+i\varepsilon) = 0, \quad (32)$$

where

$$\tilde\omega_k \approx \omega_k + \frac{\lambda_{kk}}{\mu_k\omega_k}(\sigma+\langle q\rangle^2) \quad (33)$$

is the pole of the Green's function $\langle\langle x_k|x_k\rangle\rangle^{(0)}_\omega$. From (32) it follows that

$$\varepsilon_k \approx \tilde\omega_k + \frac{1}{2\mu_k\tilde\omega_k}\mathrm{Re}\, M_k(\tilde\omega_k+i\varepsilon), \quad (34)$$

$$\Gamma_k \approx -\frac{1}{2\mu_k\tilde\omega_k}\mathrm{Im}\, M_k(\tilde\omega_k+i\varepsilon). \quad (35)$$

For our further investigations it is enough to determine the contributions to the renormalized frequencies $\varepsilon_k$ and decay coefficients $\Gamma_k$ of the cubic, $H^{(3)}_{\mathrm{int}}$, and quartet, $H^{(4)}_{\mathrm{int}}$, interactions in the first nonvanishing orders: the first in $H^{(4)}_{\mathrm{int}}$ and the second in $H^{(3)}_{\mathrm{int}}$ for $\varepsilon_k$ and the second in $H^{(4)}_{\mathrm{int}}$ for $\Gamma_k$.

Dealing with the first term in the expression (24) for the self-energy part in the same way as we did in (29), where the exact one-particle Green's functions are replaced by their lowest-order approximations (30) and (31), we arrive at an expression for the contribution to $M_k(\omega)$ of the cubic interaction in second order:

$$M^{(3)}_k(\tilde\omega_k) = 4\lambda_k^2\left[\frac{\langle q\rangle^2}{m(\tilde\omega_k^2-\Omega^2)} + \frac{\sigma}{m(\tilde\omega_k^2-4\Omega^2)}\right], \quad (36)$$

$$\varepsilon_k \equiv \omega_k + \Delta_k = \tilde\omega_k + \frac{1}{2\mu_k\tilde\omega_k}M^{(3)}_k(\tilde\omega_k). \quad (37)$$

Reasoning in a similar manner, we find an expression for the second term in (24) resulting from the quartet interaction:

$$M_k^{(4)}(\tilde{\omega}_k)$$

$$= \frac{2}{m\Omega\mu_k\omega_k\tilde{\omega}_k} \sum_{k'} \lambda_{kk'}^2 \omega_{k'}$$

$$\times \left\{ 4\langle q\rangle^2 \left[ \frac{(\Omega-\tilde{\omega}_{k'})[n(\tilde{\omega}_{k'})-n(\Omega)]}{\tilde{\omega}_k^2-(\Omega-\tilde{\omega}_{k'})^2} \right.\right.$$

$$+ \left.\frac{(\Omega+\tilde{\omega}_{k'})[1+n(\tilde{\omega}_{k'})+n(\Omega)]}{\tilde{\omega}_k^2-(\Omega+\tilde{\omega}_{k'})^2} \right]$$

$$+ \frac{2\Omega-\tilde{\omega}_{k'}}{m\Omega} \frac{[n(\Omega)+1]^2 n(\tilde{\omega}_{k'}) - n^2(\Omega)[n(\tilde{\omega}_{k'})+1]}{\tilde{\omega}_k^2-(2\Omega-\tilde{\omega}_{k'})^2}$$

$$+ \frac{2\Omega+\tilde{\omega}_{k'}}{m\Omega} \frac{[n(\Omega)+1]^2[n(\tilde{\omega}_{k'})+1] - n^2(\Omega)n(\tilde{\omega}_{k'})}{\tilde{\omega}_k^2-(2\Omega+\tilde{\omega}_{k'})^2}$$

$$+ \left. \frac{2\tilde{\omega}_{k'}}{m\Omega} \frac{n(\Omega)[n(\Omega)+1]}{\tilde{\omega}^2-\tilde{\omega}_{k'}^2} \right\}, \tag{38}$$

where $n(x)=[\exp\{x/\Theta\}-1]^{-1}$.

The decay coefficients $\Gamma_k$, which can be expressed in terms of $\operatorname{Im} M_k^{(4)}(\tilde{\omega}_k+i\varepsilon)$, are due to processes of creation (annihilation) of one or two vibrational quanta of the bistable sublattice accompanied by processes of absorption (emission) of two quanta of the matrix lattice, and also to processes of elastic scattering of the quanta of the matrix lattice that do not change the vibrational state of the bistable sublattice. On the basis of (35) and (38) we can write

$$\Gamma_k = \frac{\pi}{2\mu_k\tilde{\omega}_k m\Omega} \sum_{k'} \frac{\lambda_{kk'}^2}{\mu_{k'}\tilde{\omega}_{k'}} \left\{ 4\langle q\rangle^2 ([n(\tilde{\omega}_{k'})-n(\Omega)] \right.$$

$$\times [\delta(\tilde{\omega}_{k'}+\tilde{\omega}_k-\Omega)-\delta(\tilde{\omega}_{k'}-\tilde{\omega}_k-\Omega)] + [1$$

$$+n(\tilde{\omega}_{k'})+n(\Omega)][\delta(\tilde{\omega}_{k'}-\tilde{\omega}_k+\Omega)-\delta(\tilde{\omega}_{k'}+\tilde{\omega}_k$$

$$+\Omega)]) + \frac{1}{m\Omega}[(n(\Omega)+1)^2 n(\tilde{\omega}_{k'}) - n^2(\Omega)$$

$$\times (n(\tilde{\omega}_{k'})+1)][\delta(\tilde{\omega}_{k'}+\tilde{\omega}_k-2\Omega)-\delta(\tilde{\omega}_{k'}-\tilde{\omega}_k$$

$$-2\Omega)] + \frac{2}{m\Omega} n(\Omega)(n(\Omega)+1)[\delta(\tilde{\omega}_{k'}-\tilde{\omega}_k)$$

$$\left. - \delta(\tilde{\omega}_{k'}+\tilde{\omega}_k)] \right\}. \tag{39}$$

To do some estimates, it is enough to examine the one-dimensional model of a lattice in the Debye approximation. Plugging the expression for the coupling coefficient in the form

$$\lambda_{kk'} = \frac{\lambda}{M} \sum_{j=1}^{M} \exp\{i(k'+k)r_j\} \sqrt{\mu_k\mu_{k'}} \omega_k\omega_{k'} \tag{40}$$

($\mathbf{r}_j$ is the radius vector of the $j$th atom in the matrix lattice, and $\lambda$ is a constant whose dimensions are cm$^{-2}$) into (39), we find that for $k<k_D/2$ and $\omega_D\approx\Omega$ ($k_D$ and $\omega_D$ are the Debye wave vector and frequency) only elastic processes provide nonvanishing contributions to decay:

$$\Gamma_k \approx 4\pi\lambda^2 \left(\frac{\hbar}{2m\Omega}\right)^2 n(\Omega)[n(\Omega+1)]\omega_k \frac{k}{k_D} \frac{1}{\delta^3}, \tag{41}$$

where $\delta=\tilde{\omega}_k/\omega_k$. The above equation shows that the quartet interaction $H_{\text{int}}^{(4)}$ has almost no damping effect on vibrations in the ultrasonic frequency range. Indeed, for reasonable values of the parameters ($\lambda=8$ Å$^{-2}$, $\Omega=10^{13}$ s$^{-1}$, $T=273$ K, and $\delta=1$) we have the estimate

$$\frac{\Gamma_k}{\omega_k} \sim 10^{-2} \frac{k}{k_D}, \tag{42}$$

i.e., the decay of acoustic vibrations becomes significant only at maximum frequencies ($k\sim k_D$).

### 3.5. Temperature hysteresis of the shift and decay of the lattice mode frequencies as a result of interaction with the bistable lattice

The temperature dependence of the frequency shift $\Delta_k(T)$ is determined by the dependence of this shift on the characteristics $\langle q\rangle$, $\sigma$, and $\Omega$, which experience temperature hysteresis, and by the competition between negative (for $\Omega>\tilde{\omega}_k$) cubic and positive (for $\lambda_{kk}>0$) quartet contributions. Indeed, for instance, when the sample is cooled, the stable states of the correlated sublattice become metastable, and at the temperature $T_1$ the sublattice suddenly goes over to another stable branch (see Figs. 2 and 3), as a result of which related abrupt changes are experienced by the frequency shifts $\Delta_k$ of the acoustic modes of the matrix [see Eqs. (33), (36), and (37)] interacting with the sublattice (Fig. 4). Now, when heated, the sublattice is on this new branch up to a temperature $T_2$, after which it again suddenly returns to its old, high-temperature, stable branch, thus bringing about a sudden (discontinuous) change in $\Delta_k$ (Fig. 4). Figure 4 depicts the $\Delta_k$ vs. $T$ curves calculated by formulas (33), (36) and (37) for different ratios of the competing interactions of the third ($\lambda_k$) and fourth ($\lambda_{kk}$) orders.

The temperature dependence of the decay of an acoustic mode ($\Gamma_k$) for $k\sim k_D$ is depicted in Fig. 5 (note that here we allow for neither the nonlinearity of the matrix lattice proper, a nonlinearity that provides a nonhysteresis contribution to decay, nor for other decay mechanisms). The decay that is the largest (the high-temperature branch of curve 1) for the scattering of lattice modes by perturbations caused by above-barrier vibrations of the bistable sublattice suddenly becomes smaller (the low-temperature branch of the curve 1) when the sublattice abruptly reduces the amplitude of its vibrations after it has been ''captured'' by the global minimum as a result of cooling, The stable branch of curve 2 represents the contribution to scattering of the local minimum of the bistable potential. Actually this contribution will be smaller, since it must be multiplied by a quantity proportional to $\exp\{-\Delta F/k_B T\}$, where $\Delta F$ is the difference of free energies of the sublattice in the local and global minima; hence, the system is heated from absolute zero, the increase in decay follows almost exactly the low-temperature branch of curve 1, and then the decay suddenly increases, going over to the high-temperature branch and thus completing the hysteresis cycle.

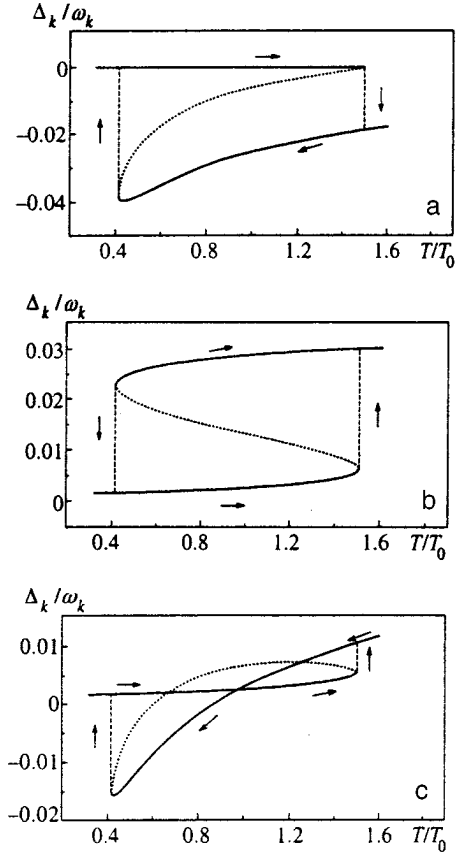

FIG. 4. Temperature dependence of the relative shift of the acoustic modes of the matrix with allowance for cubic interaction (a), quartet interaction (b), and competition of quartet and cubic interactions (c); $u_1 = 0.03$ eV, $q_1 = 0.073$ Å, $q_2 = 0.14$ Å, $T_0 = 173$ K, $\lambda_{kk}/\mu_k \omega_k^2 = 8.3$ Å$^{-2}$, and $\lambda_k = 3.74 \times 10^{-21}$ eV·Å$^{-3}$.

## 4. EXPERIMENTAL PREREQUISITES FOR THE EXISTENCE OF A BISTABLE SUBLATTICE IN SUPERCONDUCTING OXIDE CUPRATES

At present there are many papers that point to the important role that the apical oxygen atom plays in the formation of the superconducting properties of $YBa_2Cu_3O_{7-\delta}$ or compounds with a structure containing apical atoms. In the $YBa_2Cu_3O_{7-\delta}$ compound, each apical atom O(4) interacts along the crystallographic axis **c** with the two nearest neighbors, the atoms Cu(1) and Cu(2) ($l_{Cu(1)} = 1.80 - 1.86$ Å and $l_{Cu(2)} = 2.30 - 2.45$ Å; see Refs. 30 and 31), whose coordination in oxygen is not the same. The nature of the bond of the apical atom varies substantially: from covalent for the superconducting compound $YBa_2Cu_3O_{7-\delta}$ to ionic for a nonsuperconducting compound.[32] Participating in the transfer of holes from the basal planes to the $CuO_2$ planes, an apical atom manifests a number of features in the temperature dependence of the vibrational states. For instance, x-ray studies,[33–35] ion-channeling experiments,[36,37] Raman spectroscopy,[38–40] and neutron scattering measurements[41] have revealed that the total energy of an apical O(4) atom, as a function of the position along the crystallographic axis **c** has two minima. Note that pyro- and piezoelectricity have been detected in single crystals of $YBa_2Cu_3O_{7-\delta}$, which suggests that there is macroscopic polarization along the **c** axis (see, e.g., Ref. 42). The occurrence of macroscopic polarization is usually attributed to the anharmonic motion of O(4) ions.[42,43]

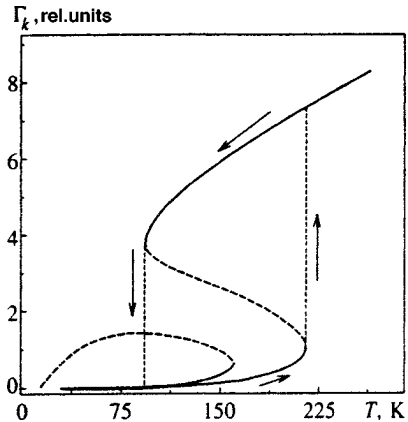

FIG. 5. Temperature dependence of the decay of the acoustic modes of the matrix in a crystal with a bistable sublattice; $u_1 = 0.03$ eV, $q_1 = 0.073$ Å, $q_2 = 0.14$ Å, and $T_0 = 173$ K.

The nontrivial dynamics of the strongly correlated apical O(4) atoms in a double-well potential must also directly manifest itself in the nature of the interaction between the vibrational states of the atoms and the electron subsystem of the crystal, which in addition to the participation of apical atoms in charge transfer from the basal plane to the $CuO_2$ plane may be one of the reasons for the formation of high-$T_c$ superconductivity.[44–46] More than that, as we will show shortly, by allowing for the interaction between the bistable oxygen sublattice and the vibrational states of the matrix lattice we can explain a number of experimentally established phenomena: the temperature hysteresis of the specific heat and thermal conductivity and of the speed and absorption of ultrasound in yttrium and bismuth cuprates.

## 5. TEMPERATURE HYSTERESIS OF THE SPEED AND DECAY COEFFICIENT OF ULTRASOUND IN HIGH-$T_c$ OXIDE CUPRATES. COMPARISON OF THEORY AND EXPERIMENT

The temperature hysteresis of the speed of ultrasound was observed by the methods of ultrasound spectroscopy[1–14] (see also the review by Lubenets et al.[15]) soon after the discovery of high-$T_c$ superconductivity in a number of oxide cuprates, including the compounds $YBa_2Cu_3O_{7-\delta}$ and $Bi_2Sr_2Ca_1Cu_2O_8$. This phenomenon can be observed not only polycrystals but also in single crystals, not only superconducting but also in nonsuperconducting high-$T_c$ compounds. The temperature interval of the hysteresis changes from sample to sample and depends on the oxygen nonstoichiometry and the way in which the sample is prepared. The values 55 and 215 K were fixed as the most reliable limits of hysteresis at the lower and higher ends of the temperature interval, although the upper limit was found to often move up to 270 K. Interestingly, Kim et al.[8] (see also Ref. 4) also observed distinct temperature hysteresis of the absorption of ultrasound in a single-crystal YBa$_2$Cu$_3$O$_{7-\delta}$ sample, and the regions of hysteresis of the speed and the absorption of ultrasound were found to coincide.

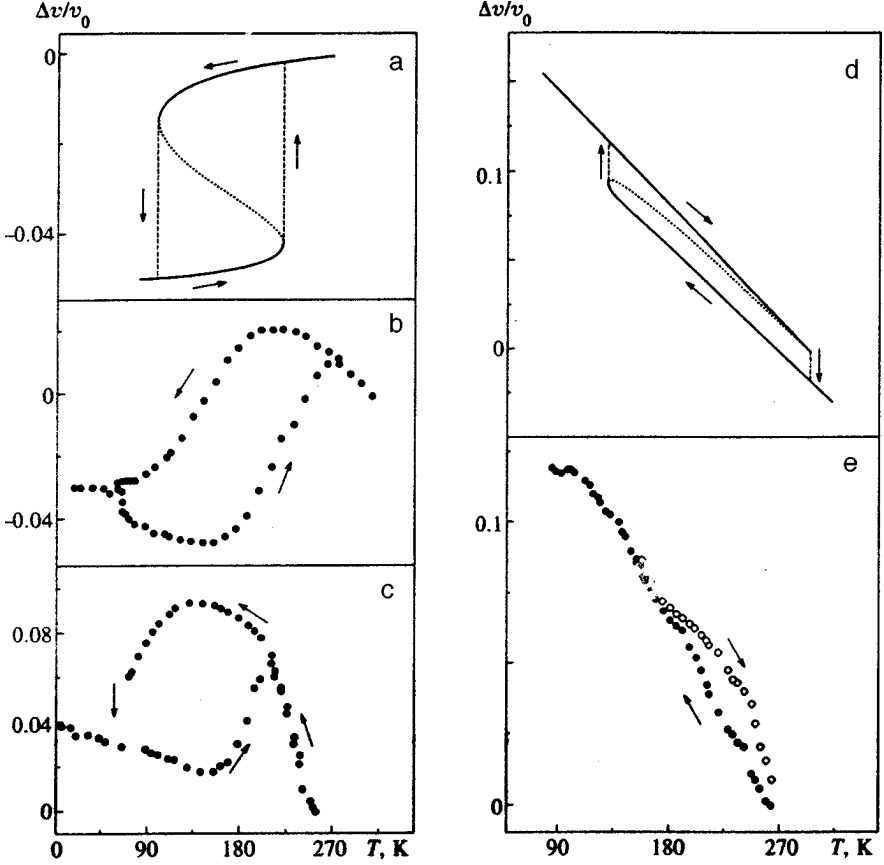

FIG. 6. Temperature dependence of the relative variation of the speed of an ultrasonic wave: (a) and (d) represent the results of calculations for a crystal with a bistable sublattice (the dotted curves represent an unstable solution), with (a) $u_1=0.03$ eV, $q_1=0.073$ Å, $q_2=0.14$ Å, and $\lambda_{kk}/\mu_k\omega_k^2=8.3$ Å$^{-2}$, and (d) $u_1=0.04$ eV, $q_1=0.073$ Å, $q_2=0.14$ Å, and $\lambda_k=3.74\times10^{-21}$ eV·Å$^{-3}$; $\Delta v/v_0=\varepsilon_k(T)/\varepsilon_k(300\text{ K})-1+(A-BT)$, with $A=0.19$ and $B=7.05\times10^{-4}$ the constants (determined from experiments[6]) of the linear dependence approximating the contribution of the main lattice. The experimental data for YBa$_2$Cu$_3$O$_7$ in the direction of the crystallographic axis **c** at 12 MHz (Ref. 2) are depicted in Fig. 6(b), and those at the frequency $1.25\times10^5$ s$^{-1}$ in Fig. 6(c). Finally, the experimental data for Bi$_2$Sr$_2$Ca$_1$Cu$_2$O$_8$ (see Ref. 6) are depicted in Fig. 6(e).

Various mechanisms for explaining these phenomena have been proposed. Among these are the redistribution of oxygen,[11,13] the motion of twinning boundaries,[47] and the presence of a ferroelectric[10] or martensitic[9] phase transition. However, no satisfactory and consistent interpretation of the temperature hysteresis of the speed and absorption of ultrasound was proposed in these papers.

We believe that a qualitative explanation of these phenomena can be found if we assume that yttrium and bismuth cuprates have a bistable oxygen sublattice that modulates the phonon spectrum of the matrix lattice. Indeed, as we showed within the scope of the general theory in Sec. 3, in this case the renormalized frequencies of the matrix and their imaginary parts acquire a hysteresis temperature dependence, so that the elastic constants of the crystal will vary in the same manner. To compare the theoretical curves with the experimental data we only need to know the empirical values of the parameters of the matrix and the sublattice and the nature of the interaction between the two.

### 5.1. Hysteresis of the speed of ultrasound in YBa$_2$Cu$_3$O$_{7-\delta}$

The temperature-dependent renormalized frequency of the long-wavelength phonons, $\varepsilon_k$, directly determines the speed $v(T)$ of an ultrasonic wave:

$$v(T)=\text{const}\cdot\varepsilon_k(T), \qquad (43)$$

where $k$ is the wave vector of the mode at whose frequency the ultrasonic measurements are carried out. As shown by further investigations, for an yttrium cuprate the quartet interaction $H_{\text{int}}^{(4)}$ in the Hamiltonian $H_{\text{int}}$ [see Eq. (11)] is the most probable one. Here, in the first-order perturbation in $H_{\text{int}}^{(4)}$ we have [see Eqs. (33) and (34)]

$$\varepsilon_k(T)\approx\omega_k[1+\lambda(\sigma+\langle q\rangle^2)], \qquad (44)$$

where $\sigma$ and $\langle q\rangle^2$ are calculated self-consistently by Eqs. (4)–(6) and we have allowed for the fact [see Eq. (40)] that $\lambda_{kk}=\lambda\mu_k\omega_k^2$. Figure 6(a) depicts the temperature dependence of the speed of ultrasound, $v(T)$, at the frequency $1.25\times10^5$ s$^{-1}$ calculated theoretically from the formulas (43), (44), and (40) with allowance for (4)–(6). As applied to YBa$_2$Cu$_3$O$_{7-\delta}$, the model parameters were specified in the following way: the "bare" frequency of an O(4) ion in the global minimum, $(\alpha/m)^{1/2}\approx 600$ cm$^{-1}$, was determined from the spectra of Raman scattering of light;[38] the position of the second, local, minimum, $q_2\approx 0.14$ Å, was found from the measured radial distribution function;[48] $q_1\approx q_2/2$; the height of the potential barrier, $u_1\approx 0.03$ eV, was chosen such that the hysteresis would land into the 100–200 K temperature range; and the coupling constant $\lambda\approx 8.3$ Å$^{-2}$ was chosen such that the theoretical values of the maximum difference in the speeds of ultrasound on the bistable branches of the hysteresis curve would agree with the experimental data. In Figs. 6(b) and 6(c) we depict, for the sake of comparison with the calculated curves, the experimental data[2,3] on the temperature dependence of the speed of longitudinal ultra-sound propagating in the ceramic and single-crystal samples of $YBa_2Cu_3O_{7-\delta}$. The theoretical curves reflect fairly well the experimentally observed behavior of the speed of ultrasound: in the hysteresis region, high speeds are realized in the cooling mode and low speeds in the heating (''thawing'') mode. Thus, at reasonable values of the model parameters, not only do the size and temperature interval of the hysteresis loop agree with the experimental data but so does the sense of tracing of the hysteresis loop in the cooling-heating cycle.

### 5.2. Hysteresis of the speed of ultrasound in $Bi_2Sr_2Ca_1Cu_2O_8$

The experimentally observed pattern of the temperature hysteresis of the speed of ultrasound in bismuth cuprates, in particular, in $Bi_2Sr_2Ca_1Cu_2O_8$ differs from that for $YBa_2Cu_3O_{7-\delta}$: the higher values of the speed of ultrasound in the hysteresis region are realized in heating, while the lower values are realized in cooling.[6,7] Such behavior of the elastic properties in bismuth cuprates can be explained if we assume that third-order anharmonicity dominates in the interaction between the matrix lattice and the bistable oxygen sublattice, so that we can assume that $\lambda_{kk}=0$. In this case, according to (34) and (36), the renormalized frequency is given by the formula

$$\varepsilon_k(T) \approx \omega_k - \frac{\lambda_k^2}{2m\Omega^2\mu_k\omega_k}(\sigma + 4\langle q \rangle^2), \quad (45)$$

where we have allowed for the fact that $\Omega \gg \omega_k$ holds in the experiment and that we have $\tilde{\omega}_k = \omega_k$ at $\lambda_{kk} \approx 0$.

Figure 6(d) depicts the temperature dependence of the speed of ultrasound at 7.5 MHz calculated by (43) and (45) with allowance for the self-consistent equations (4)–(6) for both heating and cooling. When making numerical estimates, we assumed the parameters of the bistable potential to be the same as those for yttrium compounds. We did, however, adjust the height of the potential barrier, which, like the cubic coupling constant $\lambda_k$, was chosen so that the calculated values of the temperature integral and size of hysteresis would agree best with the experimentally observed values. For the sake of comparison, in Fig. 6(e) we depict the corresponding experimental dependence[6] for ultrasonic longitudinal waves propagating in a $Bi_2Sr_2Ca_1Cu_2O_8$ single crystal. The theoretical curve represents fairly well the features of this dependence: the cooling curve lies below the heating curve, while the coincidences of the size and interval of the hysteresis in which the hysteresis loop is observed are realized at reasonable values of the parameters of the bistable and matrix lattices.

Thus, we conclude that the interaction between the metastable states of the strongly correlated oxygen sublattice (the apical O(4) atoms) and the matrix lattice in the high-$T_c$ compounds $YBa_2Cu_3O_{7-\delta}$ and $Bi_2Sr_2Ca_1Cu_2O_8$ results in renormalization of the elastic constants of the matrix lattice and, in the final analysis, an experimentally observable temperature hysteresis of the speed of ultrasound in the 60–270 K temperature range. The inversion of the hysteresis branches when yttrium cuprates are replaced by bismuth cuprates is a consequence of the change in the interaction between the bistable sublattice of O(4) atoms and the matrix.

### 5.3. Hysteresis of the absorption coefficient of ultrasound in $YBa_2Cu_3O_{7-\delta}$

As noted earlier, Kim et al.[8] clearly detected a temperature hysteresis in the absorption of a longitudinal ultrasound wave with a frequency of 5 MHz propagating in a single-crystal sample of $YBa_2Cu_3O_{7-\delta}$. A small hysteresis of the damping constant was also observed by Pal'-Val' et al.[4] in a ceramic $YBa_2Cu_3O_{7-\delta}$ at $1.25 \times 10^5 \, s^{-1}$. The absorption of ultrasound was greater when the sample was heated than when the sample was cooled. The hysteresis regions for absorption and for the speed of the wave were found to coincide, but the hysteresis loops were traced in opposite directions. The explanation of these facts follows directly from our previous discussion. Indeed, the absorption coefficient $\alpha_k$ of an ultrasound wave with a wave vector $k$ is given by the formula

$$\alpha_k = \frac{\gamma_k}{v(T)}, \quad (46)$$

where $\gamma_k$ is the attenuation of the wave, which includes as a component the decay coefficient of the hysteresis type, $\Gamma_k$, reflecting the presence of the quartet interaction $H_{int}^{(4)}$ (see Eqs. (39), (41), and (42) and Fig. 5) and the contributions from other scattering mechanisms: due to the nonlinearity of the matrix lattice proper, the effect of defects, the boundaries of the sample, etc.; $v(T)$ is the speed of ultrasound given by formulas (43) and (44). But, as noted earlier, the decay constant $\Gamma_k$ in the ultrasonic frequency range is extremely small, i.e., other scattering mechanisms are effective. Hence the anomalous temperature behavior of the absorption coefficient can be related only to the hysteresis dependence of the speed $v(T)$ of ultrasound. Figure 7(a) shows the results of a theoretical calculation of the absorption coefficient $\alpha_k$ by Eqs. (46), (43), and (44) (the attenuation $\gamma_k$ is assumed temperature-independent), and Fig. 7(c) depicts the experimental data. We see that there is not only qualitative agreement between the experimental data and the theoretical estimates (the extent and the sense of tracing of the hysteresis loop) but also a correspondence in the relative discrepancy between the absorption on the heating and cooling curves.

Note that temperature hysteresis of the speed and absorption of ultrasound of a similar type was observed by Borisov et al.[17] in $LiKSO_4$ crystals.

## 6. HYSTERESIS BEHAVIOR OF THE THERMAL PROPERTIES OF HIGH-$T_c$ OXIDE CUPRATES IN THE NORMAL STATE

### 6.1. Hysteresis behavior of specific heats

In the process of doing precision measurements, Vargas et al.[18,19] detected a temperature hysteresis of the specific heat at constant pressure, $C_P(T)$, for the high-$T_c$ cuprates $YBa_2Cu_3O_{7-\delta}$ ($\sigma=0-1$) in the 190–230 K temperature range. The heating curve was found to have a sharp peak at 220 K, while the cooling curve was found to have a fairly broad ($\sim 10\,K$) maximum at 205 K [see Fig. 8(b)]. Kumar et al.[20] obtained similar results. Vargas et al.[18,19] and Kumar et al.[20] suggest that such anomalies are due to the lattice instability; in particular, they relate the narrow peak in the heating curve to the disordering of oxygen atoms in the Cu(1)-O(1) chains, assuming that the interaction of these degrees of freedom and the lattice modes leads to a structural phase transition. We believe that these experimental facts can be explained within the scope of the idea that compounds of the form Y–Ba–Cu–O contain a strongly correlated bistable oxygen sublattice. Experiments have shown that the hysteresis interval may change by several tens of kelvins depending on oxygen content (i.e., on the way in which the sample, chiefly ceramic, is prepared). Later we will return to the problem of finding the hysteresis interval for the specific heat $C_p$ measured in the experiments.[18,19] Here we determine the contributions introduced by the bistable sublattice to the general value of the lattice specific heats at constant pressure, $C_p^{an}$, and at constant volume (constant mean displacement $\langle q \rangle$), $C_{\langle q \rangle}^{an}$. Using formulas (3) and (7), we can find the expressions for the specific heats at constant displacement and at constant pressure:

$$C_{\langle q \rangle}^{an} = k_B \left( \frac{\partial \langle H_{anh} \rangle}{\partial \Theta} \right)_{\langle q \rangle} = k_B \left( m\Omega^2 + \frac{3}{2}\gamma\sigma \right) \left( \frac{\partial \sigma}{\partial \Theta} \right)_{\langle q \rangle}, \quad (47)$$

$$p = -\left( \frac{\partial F}{\partial \langle q \rangle} \right)_\Theta = -\alpha \langle q \rangle + \beta \langle q \rangle^2 - \gamma \langle q \rangle^3 + \sigma(\beta - 3\gamma \langle q \rangle). \quad (48)$$

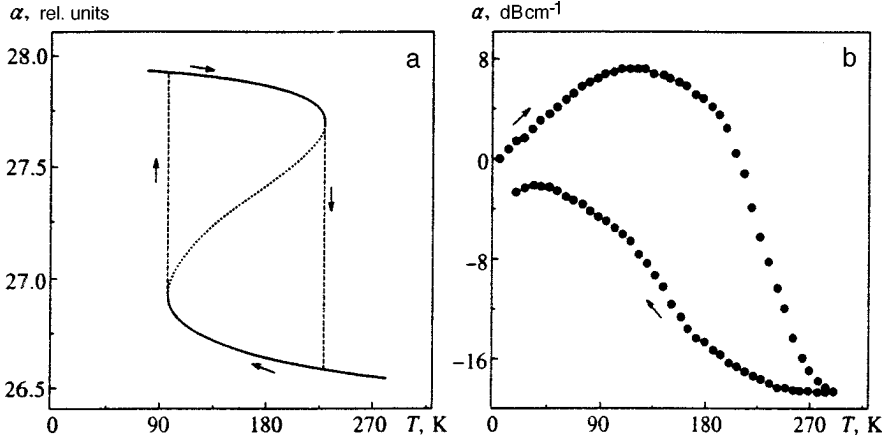

FIG. 7. Temperature dependence of ultrasound absorption: (a), the results of calculations of $\alpha \propto 1/v$ for a crystal with a bistable sublattice with $u_1 = 0.03$ eV, $q_1 = 0.073$ Å, and $q_2 = 0.14$ Å; (b), the experimentally measured absorption coefficient for the longitudinal $C_{33}$ mode at 5 MHz in the $YBa_2Cu_3O_7$ crystal.[8]

These expressions make it possible to calculate the specific heat at constant pressure:

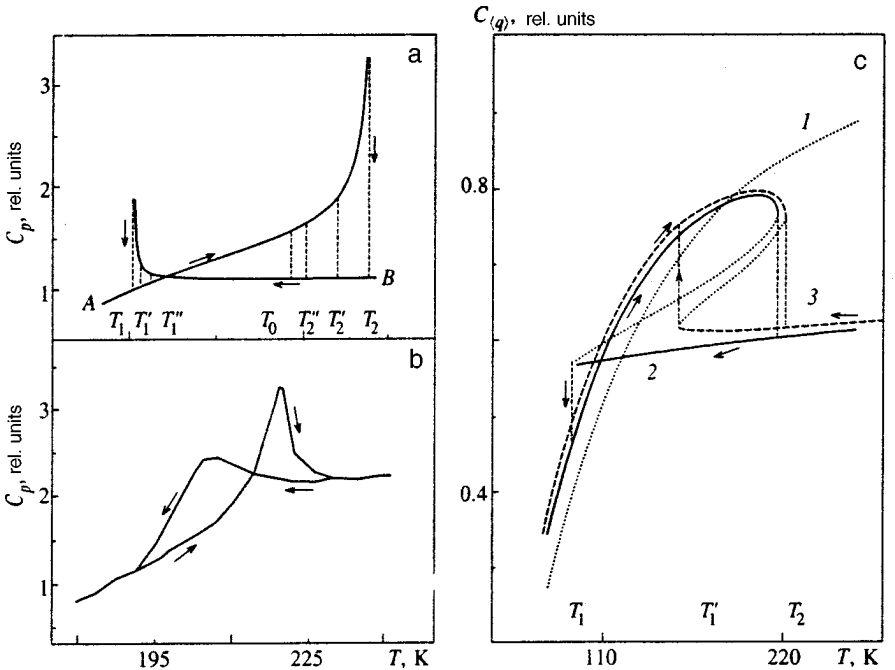

FIG. 8. Temperature dependence of the specific heat at constant pressure [(a) and (b)] and constant volume (c) for a crystal with a bistable sublattice: for a harmonic crystal (curve 1), and for a crystal with a bistable lattice (curves 2 and 3). Curves 2 and 3 were calculated for different values of the bistable potential. The dotted curves represent unstable solutions; $u_1 = 0.03$ eV, $q_1 = 0.073$ Å, and $q_2 = 0.14$ Å. Figure 8(b) depicts the experimental results for $YBa_2Cu_3O_7$ (see Ref. 19).

$$C_p^{\mathrm{an}} = C_{\langle q \rangle}^{\mathrm{an}} - k_B \frac{\Theta (\partial p/\partial \Theta)_{\langle q \rangle}^2}{(\partial p/\partial \langle q \rangle)_\Theta} = k_B \left( m\Omega^2 + \frac{3}{2}\gamma\sigma \right)$$

$$\times \left( \frac{\partial \sigma}{\partial \Theta} \right)_{\langle q \rangle} - k_B \frac{\Theta[(\beta - 3\gamma\langle q\rangle)(\partial\sigma/\partial\Theta)_{\langle q \rangle}]^2}{(\beta - 3\gamma\langle q\rangle)(\partial\sigma/\partial\langle q\rangle)_\Theta - m\Omega^2},$$
(49)

$$\left( \frac{\partial \sigma}{\partial \Theta} \right)_{\langle q \rangle} = \frac{2m\Omega^2}{\Theta} \varphi(\Omega, \sigma, \Theta),$$

$$\left( \frac{\partial \sigma}{\partial \langle q \rangle} \right)_\Theta = 2(\beta - 3\gamma\langle q\rangle)\varphi(\Omega, \sigma, \Theta). \tag{50}$$

$$\varphi(\Omega, \sigma, \Theta) = \frac{4m^2\Omega^2\sigma^2 + 4\Theta m\sigma - \hbar^2}{8\Theta^2 m^2\Omega^2 + 3\gamma(4m^2\Omega^2\sigma^2 + 4\Theta m\sigma - \hbar^2)}.$$

The temperature dependence of the specific heat at constant volume (at constant mean displacement) $C_{\langle q \rangle}^{\mathrm{an}}$, constructed from (47) with allowance for (50), exhibits temperature hysteresis [Fig. 8(c)]. The hysteresis curve is transformed according to the shape of the anharmonic potential $U$ in which the atoms O(4) move: it consists of one loop if the metastable minimum lies fairly high above the global minimum and of two loops if the metastable minimum moves downward so that the potential becomes more symmetric [see the part of the caption referring to Fig. 8(c)].

Figure 8(a) depicts the temperature dependence of the specific heat at constant pressure, $C_p^{\mathrm{an}}$, constructed from the above formulas. When the sample is cooled and the point $T_1$ is reached from the right, $C_p^{\mathrm{an}}$ becomes infinite and then suddenly drops to the finite value $A$; when the sample is heated and the point $T_2$ is reached from the left, $C_p^{\mathrm{an}}$ again becomes infinite and then suddenly drops to the value $B$. The size of the hysteresis interval $(T_1, T_2)$ depends on the values and ratios of the parameters $\alpha$, $\beta$, and $\gamma$ of the oxygen sublattice (in this specific case they were selected equal to the values in Sec. 5). In the state of thermodynamic equilibrium the system has no memory and the hysteresis disappears, i.e., the function $C_p^{\mathrm{an}}(T)$ becomes single-valued; it has, however, a singular point at the temperature $T_0$ at which the values of the free energies for the cooling and heating curves coincide. The interval $(T_0, T_1)$ in heating and the interval $(T_0, T_2)$ in cooling determine the temperature range in which the system (sublattice) passes through a sequence of alternating unstable (metastable) states.

In real compounds to which our model can be applied, it is impossible to reach the theoretical boundary points of "supercooling" ($T_1$) and "overheating" ($T_2$). The longer the system is left to itself in the region of metastable states, the higher the probability that, thanks to fluctuation processes, it will go over to the other, stable, branch of the hysteresis curve before it reaches the boundary point $T_1$ or $T_2$ and hence the narrower the hysteresis region: $T_1 \to T_1' > T_1$ and $T_2 \to T_2' < T_2$. If the lifetime of a given metastable state of the oxygen sublattice at a certain temperature $T_1''$ (the cooling curve) or $T_2''$ (the heating curve) exceeds the time the system is kept in the given state, i.e., the rate of scanning of the temperature in the experiment is such that fluctuation processes are unable to initiate the transition of the sublattice from the metastable state to a stable state, the interval $(T_1'', T_2'')$ exactly determines the real interval of hysteresis behavior of the specific heat and other properties of high-$T_c$ compounds with a bistable sublattice in conditions of the given experiment. This is the reason why in their experiments Vargas et al.[18,19] observed a hysteresis interval for the function $C_p^{\mathrm{an}}(T)$ that was narrower [Fig. 8(b)] than the theoretical interval [Fig. 8(a)]. Unfortunately, we know of no experimental data on the $C$ vs. $T$ curves for temperature scanning rates so different that the "shrinking" of the hysteresis loop can be followed as the temperature scanning rates change from high to low, i.e., as the thermodynamic parameters become more quasistatic.

### 6.2. Temperature hysteresis of the thermal conductivity in high-$T_c$ cuprates

In the process of doing precision measurements, Jezowski et al.,[21–23] Terzijska,[24] and Cohn[25] found a temperature hysteresis of the thermal conductivity of the high-$T_c$ compounds YBa$_2$Cu$_3$O$_{7-\delta}$ (1:2:3) and RBa$_2$Cu$_4$O$_8$ (1:2:4; R=Dy, Gd, and Eu) in the 70–230 K temperature range. The maximum relative discrepancy between the values of the thermal conductivity on the upper and lower branches of the hysteresis curve amounts to more than 5%. What is remarkable is that the shape of the hysteresis curve, which is single-loop for all 1:2:4 superconductors, for 1:2:3 compounds depends on the index of oxygen nonstoichiometry. At $\delta = 0$ a single loop is observed in experiments, while for oxygen-depleted ($\delta = 1$), nonconducting, compounds the hysteresis curve consists of two loops with a definite sense of tracing of the contour when the sample is first cooled and then heated. The effect of the hysteresis behavior of the thermal conductivity is unusual and interesting not only in itself but also because the reason why it appears is related to the mechanism of high-$T_c$ superconductivity.

In the high-$T_c$ compounds 1:2:3 and 1:2:4 studied in the above experiments, heat is transferred primarily by long-wavelength acoustic phonons[49] (see also our attempt in Ref. 50 to relate the anomalies in the thermal conductivity of the 1:2:3 compound to optical excitations of the sublattice of apical oxygen atoms O(4)). Below we will show that the hysteresis of the thermal conductivity can be directly related to the scattering of these acoustic modes by vibrational excitations of the bistable oxygen sublattice of the O(4) ions.

#### 6.2.1. Remarks about the formula for thermal conductivity

The Kubo formula for the kinetic thermal conductivity $K$ can be approximately expressed in terms of the square of the one-particle Green's function $G_k$ for acoustic phonons transferring heat:

$$K = \frac{k_B}{3\pi V \Theta} \sum_k \omega_k^2 v_k^2$$

$$\times \int_{-\infty}^{\infty} d\omega \frac{\exp\{\omega/\Theta\}}{(\exp\{\omega/\Theta\} - 1)^2} [\mathrm{Im}\, G_k(\omega + i\varepsilon)]^2, \tag{51}$$

where $v_k = \nabla_k \omega_k$ is the phonon group velocity, and $V$ is the volume occupied by the system. We write the Green's function as $G_k(\omega) = [\omega - \omega_k - \Delta_k(\omega) + i\gamma_k(\omega)]^{-1}$. The fre-quency shift $\Delta_k(\omega)$ of the $k$th acoustic mode is due to the interaction between this mode and the other modes of the matrix, defects, electrons, etc., and the bistable sublattice. The same can be said of the rate of scattering of an acoustic phonon, $\gamma_k$, related to the lifetime of this phonon by the formula $\tau_k(\omega) = [2\gamma_k(\omega)]^{-1}$. We write the Green's function approximately as $G_k(\omega + i\varepsilon) \approx (\omega - \omega_k - \Delta_k + i\gamma_k)^{-1}$, replacing $\omega$ in the expressions for the frequency shift and decay by $\omega_k$: $\Delta_k = \Delta_k(\omega_k)$ and $\gamma_k = \gamma_k(\omega_k)$.

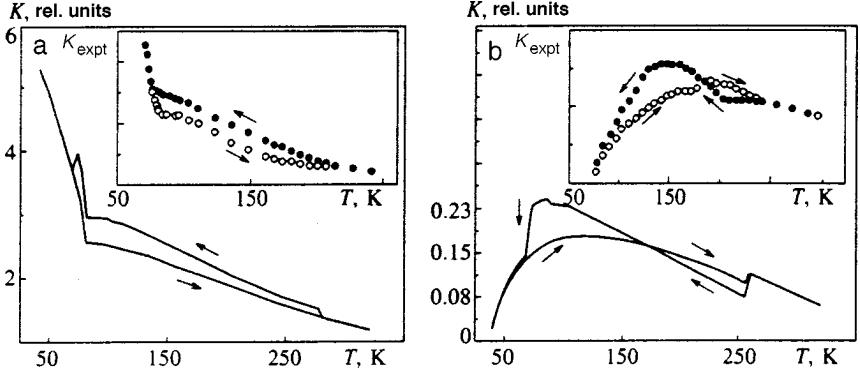

FIG. 9. Theoretical curves representing the temperature dependence of specific heat in crystals with a bistable sublattice: (a), cubic interaction between the bistable sublattice and the matrix is dominant; (b), competition between cubic and quartet interactions; $u_1 = 0.03$ eV, $q_1 = 0.073$ Å, $q_2 = 0.14$ Å, $T_0 = 173$ K, $\lambda_{kk}/\mu_k \omega_k^2 = 8.3$ Å$^{-2}$, and $\lambda_k = 3.74 \times 10^{-21}$ eV·Å$^{-3}$. The insets schematically depict the experimental results[21–23] for RBa$_2$Cu$_4$O$_8$ and YBa$_2$Cu$_3$O$_7$ (a) and RBa$_2$Cu$_3$O$_6$ (b).

Then for $\gamma_k/\omega_k \ll 1$ we can write the expression (51) in the form[51]

$$K = \frac{k_B}{3\Theta^2 V} \sum_k \omega_k^2 v_k^2 \frac{\exp\{\varepsilon_k/\Theta\}}{(\exp\{\varepsilon_k/\Theta\} - 1)^2} \frac{1}{2\gamma_k}, \qquad (52)$$

where $\varepsilon_k = \omega_k + \Delta_k$. This formula differs from the standard expression[52] in explicitly allowing for the effect of the frequency shift $\Delta_k$. Below we assume that the frequency shift $\Delta_k = \Delta_k(T)$ is caused solely by the interaction between the acoustic phonons and the vibrations of the bistable oxygen sublattice, with the frequency renormalization due to the nonlinearity of the matrix lattice proper, the interaction with defects and charge carriers, etc., included in the definition of $\omega_k$. Such separation of contributions is natural since they differ qualitatively: as established earlier, the contribution of the bistable sublattice to the renormalization of the frequencies of the matrix lattice is of a hysteresis nature, i.e., its temperature behavior when the sample is cooled differs from its behavior when the sample undergoes heating (''thawing'').

The formula that is commonly used[53] to calculate the thermal conductivity of the high-$T_c$ 1:2:3 and 1:2:4 compounds must be modified; into the exponents we must introduce the frequency shift $\Delta_k(T)$ caused by the heat-transferring phonons scattered by the bistable oxygen sublattice, which means that in the reciprocal phonon lifetime, $\tau_k^{-1}$, we must separate this additional relaxation channel, which leads to decay, earlier denoted by $\Gamma_k$. As a result we have a formula for numerical calculations:

$$K = \text{const} \cdot T^3 \int_0^{\omega_{\max}/k_B T} dx \, x^4 \exp\left\{ x + \frac{\Delta(x,T)}{k_B T} \right\}$$

$$\times \left( \exp\left\{ x + \frac{\Delta(x,T)}{k_B T} \right\} - 1 \right)^{-2} \tau(x,T), \qquad (53)$$

where

$$\tau^{-1} = \tau_b^{-1} + \tau_d^{-1} + \tau_e^{-1} + \tau_{mp}^{-1} + \tau_{bs}^{-1} = A + BT^4 x^4$$
$$+ CTxg(x,T/T_c) + Ex^2T^3 + D\Gamma(x,T). \qquad (54)$$

The terms in the expression for the relaxation time describe the scattering by the boundaries, point defects, electrons, matrix phonons, and the bistable sublattice, and $g(x,T/T_c)$ is the ratio of the relaxation times in the normal and superconducting states;[53] the functions $\Delta(x,T)$ and $\Gamma(x,t)$ are specified by Eqs. (37), (33), (36), (39), and (41), with $x \equiv \omega_k/k_B T$. When necessary, we can allow for the contribution to second order in $H_{\text{int}}^{(4)}$ in the frequency shift by using Eqs. (34) and (38). The effect of $\Gamma(x,t)$ is not appreciable against the background of other mechanisms, but it is obvious that it is the frequency shift $\Delta(x,t)$ of the acoustic modes, which enters into the exponential factors in (53), that basically determines the temperature dependence of $K(T)$.

### 6.2.2. Numerical estimates of the thermal conductivity

Numerical calculations of $K(T)$ by formula (53) with allowance for the expressions (37), (33), (36), (39), (41), and (54) and for the self-consistent equations (4)–(6) show that when the cubic interaction between the matrix and the bistable sublattice dominates, the hysteresis curve for the thermal conductivity in the 70–230 K temperature range consists of a single loop [Fig. 9(a)]. It is this hysteresis loop that is observed in experiments involving stoichiometric ($\delta = 0$) samples of 1:2:3 and 1:2:4 compounds [see the inset in Fig. 9(a)]. What is interesting is that when the quartet interaction dominates ($\lambda_{kk} > 0$), the temperature dependence of $K(T)$ basically retains its shape, although the hysteresis cycle is traced in the opposite direction, with the sense of tracing of the cycle agreeing with the one observed in experiments if $\lambda_{kk} < 0$. As a result of the competition of comparable contributions of the cubic and quartet interactions, which probably occurs for nonconducting ($7 - \delta = 6$) samples of 1:2:3 compounds, the hysteresis part of the $K$ vs. $T$ curves becomes two-loop [Fig. 9(b)] and the temperature regions of the two-loop and the one-loop hysteresis coincide, as they do in experiments.

Thus, the interaction between the metastable states of the sublattice of apical O(4) atoms and the matrix gives rise to a temperature hysteresis of the renormalized frequencies of the heat-transferring acoustic modes, which in the final analysis is the reason for the hysteresis behavior of the thermal conductivity in the high-$T_c$ 1:2:3 and 1:2:4 compounds; here the region of bistable thermal conductivity coincides with the region of bistability of the oxygen sublattice. The form of the hysteresis curve, one-loop or two-loop, and the sense of tracing of the hysteresis loops for $K(T)$ depend on the ratio of the contributions, to the renormalization, of the cubic and quartet terms in the interaction between the matrix and the bistable sublattice.

## 7. CONCLUSION

We have examined the situation in which in a crystal lattice with a multiatomic basis the atoms of a certain species perform optical vibrations in an asymmetric double-well potential generated by the field of the matrix lattice. If the motion of such atoms is strongly correlated, i.e., is of a cooperative nature, this suppresses fluctuation above-barrier transitions of separate atoms from one energy minimum to other, in view of which (and because of broken symmetry of the potential) metastable states may become realizable for the ensemble of atoms considered, thus producing a bistable sublattice. The critical temperature of the transition of such a sublattice from vibrations inside the global minimum to above-barrier vibrations under heating does not coincide with the temperature of the inverse transition from above-barrier vibrations to intrawell vibrations in the inverse process of cooling. Here the temperature hysteresis is characteristic both for the dynamic parameters of the sublattice and for the statistical-thermodynamic parameters of the sublattice, and the size of the hysteresis interval and its position on the temperature axis are determined primarily by the energy difference of the global and local minima and the height of the potential barrier. Due to the nonlinear interaction between the metastable states of the sublattice and the vibrational states of the matrix lattice, a hysteresis temperature dependence becomes a characteristic feature of the renormalized frequencies of the lattice modes and the decay coefficients for these modes, which in the final analysis gives rise to hysteresis in the elastic and thermal properties of crystals. The ideas developed in this paper result in an interpretation (which agrees fairly well with the experimental data) of the hysteresis temperature behavior of the speed and absorption of ultrasound and of the specific heat and thermal conductivity in superconducting yttrium and bismuth cuprates.

The analysis has shown that in bismuth cuprates the cubic interaction between the degrees of freedom of the bistable sublattice and the matrix lattice is realized, while in yttrium cuprates the interaction is of the quartet type. In accordance with this it occurs that the higher values of the speed of ultrasound are observed in bismuth cuprates under heating, while the lower values are observed under cooling; the situation is the opposite in yttrium cuprates.

Our estimates made it possible to establish that the hysteresis of the absorption of ultrasound is due entirely to the hysteresis dependence of the speed of ultrasound. Hence, knowing the nature of the behavior of the latter, we can predict when the absorption is higher: under heating or under cooling.

While the bistability of the specific heat of the crystal is related to the anharmonic contribution of the bistable sublattice to the total specific heat and the shape of the hysteresis, one-loop or two-loop, depends on the relative positions of the global and local minima of the potential, the reason for the hysteresis of the thermal conductivity is the hysteresis temperature renormalization of the heat-transferring acoustic modes and the shape of the bistable thermal-conductivity curve (one-loop or two-loop), and the sense in which the loops are traced depends on the ratio of the contributions, to the renormalization, of the cubic and quartet terms in the interaction between the matrix and the bistable sublattice.

In all the effects examined in this paper, the region of temperature hysteresis coincides with that of the hysteresis of the bistable sublattice, since the elastic and thermal properties of the crystal depend on the main parameters of the sublattice: its statistical-mean displacement, the displacement variance, and the effective vibration frequency.

### ACKNOWLEDGMENTS

This research was supported by the Belorussian Fund for Basic Research (Grant Φ96-342).

*)E-mail: saiko@physics.by

[1] S. Ewert, S Guo, P. Lemmens *et al.*, Solid State Commun. **64**, 1153 (1987).
[2] M.-F. Xu, A. Schenstrom, Y. Hong *et al.*, IEEE Trans. Magn. **25**, 2414 (1989).
[3] V. D. Natsik, P. P. Pal'-Val', J. Engert, and H.-J. Kaufmann, Fiz. Nizk. Temp. **15**, 836 (1989) [Sov. J. Low Temp. **15**, 463 (1989)].
[4] L. N. Pal'-Val', V. I. Dotsenko, P. P. Pal'-Val', V. D. Natsik, I. F. Kislyak, and A. A. Shevchenok, Sverkhprovodimost': Fiz., Khim., Tekhnol. **3**, 1244 (1990) [Supercond., Phys. Chem. Technol. **3**(6), S92 (1990)].
[5] Z. Zhao, S. Adenwalla, A. Moreau, and J. B. Ketterson, J. Less-Common Met. **149**, 451 (1989).
[6] Y.-N. Wang, J. Wu, H.-M. Shen, J.-S. Zhu, X.-H. Chen, Y.-F. Yan, and Z.-X. Zhao, Phys. Rev. B **41**, 8981 (1990).
[7] X.-D. Xiang, M. Chung, J. W. Brill *et al.*, Solid State Commun. **69**, 833 (1989).
[8] T. J. Kim, J. Kowalewski, W. Assmus, and W. Grill, Z. Phys. B **78**, 207 (1990).
[9] X. Chen, Y. Wang, and H. Shen, Phys. Status Solidi A **113**, K85 (1989).
[10] V. Müller, C. Hucho, K. deGroot *et al.*, Solid State Commun. **72**, 997 (1989).
[11] L. G. Mamsurova, K. S. Pigal'skiĭ, V. P. Sakun, A. I. Shushin, and L. G. Shcherbakova, Zh. Éksp. Teor. Fiz. **98**, 978 (1990) [Sov. Phys. JETP **71**, 544 (1990)].
[12] E. Biagi, E. Borchi, R. Garre *et al.*, Phys. Status Solidi A **138**, 249 (1993).
[13] T. Fukami, R. Kondo, T. Kobayashi *et al.*, Physica C **241**, 336 (1995).
[14] P. P. Pal-Val, L. N. Pal-Val, V. D. Natsik, I. F. Kislyak, and V. I. Dotsenko, Fiz. Nizk. Temp. **22**, 1452 (1996) [Low Temp. Phys. **22**, 1103 (1996)].
[15] S. V. Lubenets, V. D. Natsik, and L. S. Fomenko, Fiz. Nizk. Temp. **21**, 475 (1995) [Low Temp. Phys. **21**, 367 (1995)].
[16] L. Godfrey and J. Philip, Phys. Rev. B **54**, 15708 (1996).
[17] B. F. Borisov, E. V. Charnaya, and M. Ya. Vinogradova, Phys. Status Solidi B **199**, 51 (1997).
[18] R. A. Vargas, M. Chacon, J. C. Trochez, and I. Palacoix, Phys. Lett. A **139**, 81 (1989).
[19] R. A. Vargas, P. Prietto, M. Chacon, and J. C. Trochez, Phys. Lett. A **152**, 105 (1991).
[20] S. Kumar, S. P. Pai, R. Pinto, and D. Kumar, Physica C **215**, 286 (1993).
[21] A. Jezowski, J. Klamut, R. Horyn, and K. Rogacki, Supercond. Sci. Technol. **1**, 296 (1989).
[22] A. Jezowski, Solid State Commun. **71**, 419 (1989).
[23] A. Jezowski, J. Klamut, and B. Dabrowski, Phys. Rev. B **52**, R7030 (1995).

[24] B. M. Terzijska, Cryogenics **32**, 60 (1992).
[25] J. L. Cohn, J. Supercond. **8**, 457 (1995).
[26] V. Kh. Kozlovskiĭ, Zh. Éksp. Teor. Fiz. **30**, 766 (1956) [Sov. Phys. JETP **3**, 601 (1956)].
[27] Yu. N. Gornostyrev, M. I. Katsnel'son, and A. V. Trefilov, JETP Lett. **56**, 529 (1992).
[28] N. M. Plakida, in *Statistical Physics and Quantum Field Theory* [in Russian], N. N. Bogolyubov (Ed.), Nauka, Moscow (1973).
[29] D. N. Zubarev, *Nonequilibrium Statistical Thermodynamics*, Consultants Bureau, New York (1974).
[30] J. D. Jorgensen, B. W. Veal, A. P. Paulikas, L. J. Nowicki, G. W. Crabtree, H. Claus, and W. K. Kwok, Phys. Rev. B **41**, 1863 (1990); **42**, 995 (1990).
[31] R. J. Cava, A. W. Hewat, E. A. Hewat *et al.*, Physica C **165**, 419 (1990).
[32] V. E. Gusakov, Fiz. Nizk. Temp. **21**, 805 (1995) [Low Temp. Phys. **21**, 621 (1995)]; Ya. S. Bobovich, Usp. Fiz. Nauk **167**, 973 (1997) [Phys. Usp. **40**, 925 (1997)].
[33] D. Conradson, I. Raistrick, and A. R. Bishop, Science **248**, 1394 (1990).
[34] De Leon J. Mustre, S. D. Conradson, I. Batistić, and A. R. Bishop, Phys. Rev. Lett. **65**, 1675 (1990).
[35] E. A. Stern, M. M. Qian, Y. Y. Yacoby *et al.*, Physica C **209**, 331 (1993).
[36] J. Remmel, O. Meyer, J. Geerk, J. Reiner, and G. Linker, A. Erb and G. Müller-Vogt, Phys. Rev. B **48**, 16168 (1993).
[37] R. P. Sharma, L. E. Rehn, P. M. Baldo, and J. Z. Lin, Phys. Rev. Lett. **62**, 2869 (1989).
[38] D. Mihailovic, K. F. McCarty, and D. S. Ginley, Phys. Rev. B **47**, 8910 (1993).
[39] G. Ruani, C. Taliani, M. Muccini *et al.*, Physica C **226**, 101 (1994).
[40] L. V. Gasparov, V. D. Kulakovskii, V. B. Timofeev, and E. Sherma, J. Supercond. **8**, 27 (1995).
[41] H. A. Mook, B. C. Chakoumakos, M. Mostoller, A. T. Boothroyd, and D. McK. Paul, Phys. Rev. Lett. **69**, 2272 (1992).
[42] D. Mihailovic, I. Poberaj, and A. Mertelj, Phys. Rev. B **48**, 16634 (1993).
[43] T. Galbaatar, N. M. Plakida, and S. L. Drechsler, Preprint JINR E17-94-299, Dubna (1994).
[44] R. Aoki, H. Murakami, and T. Kita, Physica C **235–240**, 1891 (1994).
[45] A. P. Saĭko, V. E. Gusakov, and V. S. Kuz'min, JETP Lett. **56**, 411 (1992).
[46] A. P. Saĭko and V. E. Gusakov, Zh. Éksp. Teor. Fiz. **108**, 757 (1995) [JETP **81**, 413 (1995)].
[47] J. F. Smith and D. Wohlleben, Z. Phys. B **72**, 323 (1988).
[48] J. Ranninger, Z. Phys. B **84**, 167 (1991).
[49] B. M. Anderson and B. Sundqvist, Phys. Rev. B **48**, 3575 (1993).
[50] A. P. Saĭko, V. E. Gusakov, and V. S. Kuz'min, JETP Lett. **57**, 116 (1993).
[51] B. Deo and S. N. Behera, Phys. Rev. **141**, 738 (1966).
[52] P. Carruthers, Rev. Mod. Phys. **33**, 92 (1961).
[53] B. M. Anderson, B. Sundqvist, J. Niska, and B. Loberg, Phys. Rev. B **49**, 4189 (1994).